\documentstyle[epsfig]{aipproc}

\def\cpt{\chi -PT}
\def\lsc{\Lambda _\chi}

\def\be{\begin{equation}}
\def\ee{\end{equation}}
\def\bea{\begin{eqnarray}}
\def\eea{\end{eqnarray}}

\def\lsc{\Lambda _\chi}
\def\cpt{\chi PT}

\def\vpee{{\vec p}}

\begin{document}
\title{Elusive Neutrons from Nuclei\\in Effective Field Theory}

\author{Silas R.~Beane$^*$}
\address{$^*$Department of Physics, University of Maryland\\
College Park, MD 20742-4111}

\maketitle

\begin{abstract}
We review recent computations of neutral pion photoproduction and
Compton scattering on the deuteron in baryon chiral perturbation
theory. Progress in extracting the neutron electric dipole amplitude,
which is relevant in neutral pion photoproduction, and the neutron
polarizabilities, which are relevant in Compton scattering, is
discussed.
\end{abstract}

\section*{Introduction}

\noindent The absence of suitable neutron targets in low-energy
scattering experiments requires the use of nuclear targets like
deuterium and helium in order to extract neutron scattering data.  The
extent to which neutron data can be reliably extracted depends on how
under control the errors are in computing the nuclear corrections to
free nucleon motion.  Of course precise calculations of hadron
processes are possible only where a small dimensionless expansion
parameter is identified. This is the main motivation behind the
ongoing intense effort to develop a perturbative theory of nuclear
interactions~\cite{monster}.  The dimensionless parameters relevant to
low energy QCD and therefore to nuclear physics consist of ratios of
external momenta to various characteristic energy scales, like the
nucleon mass.  Effective field theory is the technology which develops
a hierarchy of scales into a perturbative expansion of physical
observables.

In this paper we describe several recent effective field theory
calculations whose objective is to extract neutron properties from
nuclear scattering processes in a systematic way. We first discuss a
computation of neutral pion photoproduction on the deuteron and its
dependence on nucleon parameters.  We then describe a calculation of
Compton scattering on the deuteron at photon energies of order the
pion mass. Here the ultimate objective is to learn about neutron
polarizabilities from nuclear Compton scattering. The basic
power-counting scheme is reviewed in the first section. In the second
section we discuss photoproduction on the deuteron.  The third section
is dedicated to Compton scattering on the deuteron.

\section*{Weinberg power-counting}
\label{sec:power}

\noindent At energies well below the chiral symmetry breaking scale,
$\lsc\sim 4 \pi f_\pi \sim {M} \sim {m_\rho}$, the interactions of
pions, photons and nucleons can be described systematically using an
effective field theory.  This effective field theory, known as chiral
perturbation theory ($\cpt \,$), reflects the observed QCD pattern of
symmetry breaking.  In QCD the chiral $SU(2)_L \times SU(2)_R$
symmetry is spontaneously broken.  Here we are interested in processes
where the typical momenta of all external particles is $p\ll\lsc$, so
we identify our expansion parameter as $p/\lsc$.  In QCD $SU(2)_L
\times SU(2)_R$ is softly broken by the small quark masses. This
explicit breaking implies that the pion has a small mass in the
low-energy theory.  Since ${m_\pi}/\lsc$ is then also a small
parameter, we have a dual expansion in $p/\lsc$ and ${m_\pi}/\lsc$. We
take $Q$ to represent either a small momentum {\it or} a pion mass.

In few-nucleon systems, a complication arises in $\cpt$ due to the
existence of shallow nuclear bound states and related infrared
singularities in $A$-nucleon reducible Feynman diagrams evaluated in
the static approximation~\cite{weinnp}. The fundamental problem is
that nuclear physics introduces a new mass scale, the nuclear binding
energy, which is very small compared to a typical hadronic scale like
$\lsc$.  One way to overcome this difficulty is to adopt a modified
power-counting scheme in which $\cpt$ is used to calculate an
effective potential which generally consists of all $A$-nucleon
irreducible graphs. The $S$-matrix, which includes all reducible
graphs as well, is then obtained through iteration by solving a
Lippmann-Schwinger equation~\cite{weinnp}. This version of nuclear
effective theory is known as the Weinberg formulation.  There now
exists a competing power-counting scheme in which all nonperturbative
physics responsible for the presence of low-lying bound states arises
from the iteration of a single operator in the effective theory, while
all other effects, including all higher dimensional operators {\it
and} pion exchange, are treated perturbatively~\cite{pc,ksw}. This
version of the effective theory is known as the Kaplan-Savage-Wise
(KSW) formulation. This is relevant here because Compton scattering on
the deuteron has been computed to next-to-leading order in the KSW
formulation~\cite{martinetal}.  We will discuss this result and its
relation to our calculation.  A comprehensive and up-to-date review of
nuclear applications of effective field theories can be found in
Ref.~\cite{monster}.

It should be noted that typical nucleon momenta inside the deuteron
are small---on the order of $\sqrt{MB}$ or $m_\pi$, with $B$ the
deuteron binding energy---and consequently, {\it a priori} we expect
no convergence problems in the $\cpt$ expansion of any low-momentum
electromagnetic or pionic probe of the deuteron.  Although in
principle we could use wavefunctions computed in $\cpt$, we will
consider wavefunctions generated using modern nucleon-nucleon
potentials~\cite{wein1}. Generally we find that any wavefunction with
the correct binding energy gives equivalent results to within the
theoretical error expected from neglected higher orders in the chiral
expansion.  Presumably we are insensitive to short distance components
of the wavefunction because we are working at low energies and the
deuteron is a large object.

\section*{Photoproduction on the deuteron}
\label{sec:app}

\begin{figure}[!] 
\centerline{\epsfig{file=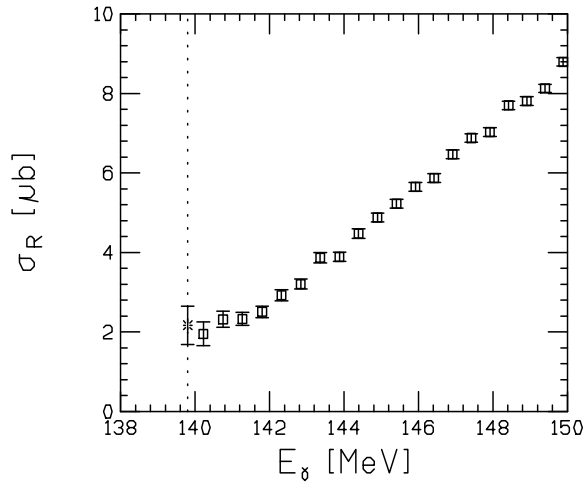,height=2.5in,width=3in}}
\vspace{10pt}
\caption{SAL data versus $\cpt$ prediction (star).}
\label{fig1}
\end{figure}

\noindent A striking example of the power of effective field theory
is neutral pion photoproduction on the deuteron at threshold. The
$O(Q^4)$ $\cpt$ prediction for the electric dipole amplitude in
neutral pion photoproduction on the deuteron is~\cite{silas2}

\begin{eqnarray} E_d &=& E_d^{ss} + E_d^{tb,3} + E_d^{tb,4}\nonumber
\\ &=& (0.36 - 1.90 - 0.25) \times 10^{-3}/m_{\pi^+} \nonumber \\ &=&
(-1.8 \pm 0.2) \times 10^{-3}/m_{\pi^+} 
\end{eqnarray} 
where $E_d^{tb,3}$ and $E_d^{tb,4}$ represent $O(Q^3)$ and $O(Q^4)$
three-body corrections, respectively, and

\begin{eqnarray} 
{E^{ss}_{d}} &=&
\frac{1+{m_\pi}/{M}} {1+{m_\pi}/{M_d}} \, \biggl\{ \frac{1}{2} \,
({E_{0+}^{{\pi ^0}p}}+{E_{0+}^{{\pi ^0}n}} ) \, \int d^{3}{p} \; 
{{\phi}}^{\ast}_{f}(\vec{p}) \, \vec{\epsilon} \cdot \vec{J} \,
{{\phi}}_{i}(\vec{p}-\vec{k}/2) \nonumber \\
&-& \frac{k}{M} \, \hat{k} \cdot \int d^{3}{p} \; \hat{p} \,
\frac{1}{2} \, ({P_{1}^{{\pi ^0}p}}+{P_{1}^{{\pi ^0}n}} ) \,
{{\phi}}^{\ast}_{f}(\vec{p})\, \vec{\epsilon} \cdot \vec{J} \,
{{\phi}}_{i}(\vec{p}-\vec{k}/2) \biggr\}, \,\, 
\end{eqnarray}
where $\phi$ represents the deuteron wavefunction, here taken from the
Argonne V18 potential, $M_d$ is the deuteron mass and $\vec k$ is the
photon momentum. The $O(Q^4)$ $\cpt$ values for the electric
dipole amplitudes of the proton and neutron are~\cite{cptpreds}

\begin{equation} E_{0+}^{{\pi ^0}p} = -1.16 \times 10^{-3}/m_{\pi^+}
\, \qquad E_{0+}^{{\pi ^0}n} = +2.13 \times 10^{-3}/m_{\pi^+}. \,
\end{equation} Note the large value of the neutron electric dipole
amplitude. The corresponding neutron cross section is a factor of four
larger than the proton cross section, in complete violation of
classical intuition. (These are not bona fide predictions since they
involve counterterms determined by resonance saturation.) The proton
empirical value from MAMI~\cite{mami} is

\begin{equation}
E_{0+}^{{\pi ^0}p} = (-1.31 \pm 0.08) \times 10^{-3}/m_{\pi^+},
\end{equation}
in agreement with the $\cpt$ prediction. In order to test
the neutron prediction we must consider the deuteron.  The SAL
empirical value~\cite{sally} for the deuteron is

\begin{eqnarray}
E^{exp}_d &=& (-1.45 \pm 0.09)  \times 10^{-3}/m_{\pi^+} 
\end{eqnarray}
and therefore overlaps with the $\cpt$ prediction within 1.5
$\sigma$. One may wonder about the sensitivity of the deuteron
electric dipole amplitude to the neutron contribution. For instance,
if we set $E_{0+}^{{\pi ^0}n} = 0$ the $\cpt$ prediction becomes $E_d
= -2.6$, completely at odds with the experimental value. This result
is a striking confirmation of the large $\cpt$ prediction for the
neutron. It is interesting that models miss the large chiral loop
effects and therefore predict a much smaller neutron electric dipole
amplitude. See Fig. \ref{fig1}.

\section*{Deuteron Compton scattering}
\label{sec:entre}

\noindent Nucleon Compton scattering has been studied in $\cpt$ in
Ref.~\cite{ulf1}, where the following results for the polarizabilities
were obtained to order $Q^3$:

\begin{eqnarray}
\alpha_p=\alpha_n=\frac{5 e^2 g_A^2}{384 \pi^2 f_\pi^2 m_\pi} &=&12.2 \times
10^{-4} \, {\rm fm}^3; \label{eq:alphaOQ3}\nonumber\\ 
\beta_p=\beta_n=\frac{e^2 g_A^2}{768 \pi^2 f_\pi^2 m_\pi}&=& 1.2 \times 
10^{-4} \, {\rm fm}^3. \label{eq:betaOQ3}
\end{eqnarray}
Here we have used $g_A=1.26$ for the axial coupling of the nucleon,
and $f_\pi=93$ MeV as the pion decay constant. Note that the
polarizabilities are {\it predictions} of $\cpt$ at this order.  The
$O(Q^3)$ $\cpt$ predictions diverge in the chiral limit because they
arise from pion loop effects.

Recent experimental values for the proton polarizabilities are
\cite{newanalysis} \footnote{These are the result of a model-dependent
fit to data from Compton scattering on the proton at several angles
and at energies ranging from 33 to 309 MeV.}

\begin{eqnarray}
\alpha_p + \beta_p=13.23 \pm 0.86^{+0.20}_{-0.49} \times 10^{-4} \, {\rm fm}^3,
\nonumber\\
\alpha_p - \beta_p=10.11 \pm 1.74^{+1.22}_{-0.86} \times 10^{-4} \, {\rm fm}^3,
\label{polexpt}
\end{eqnarray}
where the first error is a combined statistical and systematic error,
and the second set of errors comes from the theoretical model
employed. These values are in good agreement with the $\cpt$
predictions.

On the other hand, the neutron polarizabilities are difficult to
obtain experimentally and so the corresponding $\cpt$ prediction is
not well tested.  One way to extract neutron polarizabilities is to
consider Compton scattering on nuclear targets.  Consider coherent
photon scattering on the deuteron. The cross section in the forward
direction naively goes as:

\begin{equation}
\left.\frac{d \sigma}{d \Omega} \right|_{\theta=0}
\sim (f_{Th} - (\alpha_p + \alpha_n) \omega^2)^2.
\end{equation} 
The sum $\alpha_p + \alpha_n$ may then be accessible via its
interference with the dominant Thomson term for the proton,
$f_{Th}$~\cite{hornidge}. This means that with experimental knowledge
of the proton polarizabilities it may be possible to extract those for
the neutron.  Coherent Compton scattering on a deuteron target has
been measured at $E_\gamma=$ 49 and 69 MeV by the Illinois group
\cite{lucas}.  An experiment with tagged photons in the energy range
$E_\gamma= 84.2-104.5$ MeV is under analysis at Saskatoon \cite{SAL}.

Clearly the amplitude for Compton scattering on the deuteron involves
mechanisms other than Compton scattering on the individual constituent
nucleons. Hence, extraction of nucleon polarizabilities requires a
theoretical calculation of Compton scattering on the deuteron that is
under control in the sense that it accounts for {\it all} mechanisms
to a given order in a systematic expansion in a small parameter.

In the remainder of this paper we will review a recent computation of
Compton scattering on the deuteron for incoming photon energies of
order 100 MeV in the Weinberg formulation~\cite{silas3}. As in the
computation of the electric dipole amplitude in photoproduction,
baryon $\cpt$ is used to compute an irreducible scattering kernel
(here to order $Q^3$) which is then sewn to external deuteron
wavefunctions.

\begin{figure}[t,h,b,p]
\qquad\qquad\qquad\psfig{figure=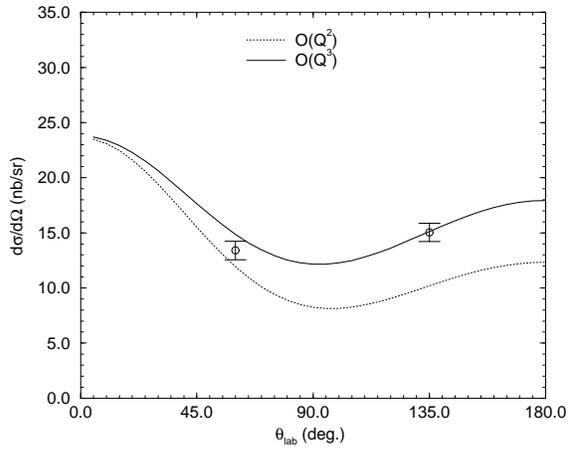,height=2.7in}
\caption{Results of the $O(Q^2)$ (dotted line) and $O(Q^3)$ (solid
    line) calculations at a photon laboratory energy of 69 MeV.
\label{fig:fig9}}
\end{figure}

\begin{figure}[t,h,b,p]
\qquad\qquad\qquad\psfig{figure=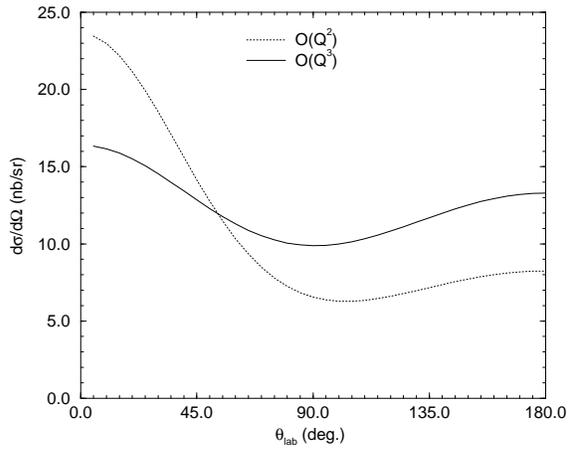,height=2.7in}
\caption{Results of the $O(Q^2)$ (dotted line) and $O(Q^3)$ (solid
    line) calculations at a photon laboratory energy of 95 MeV.
\label{fig:fig10}}
\end{figure}

In Figures~\ref{fig:fig9} and \ref{fig:fig10} we display our results
at 69 and 95 MeV. For comparison we have included the calculation at
$O(Q^2)$ where
the $\gamma N$ $T$-matrix in the single-scattering
contribution is given by the Thomson term on a single nucleon. It is
remarkable that to $O(Q^3)$ no unknown counterterms appear.  All
contributions to the kernel are fixed in terms of known pion and
nucleon parameters such as $m_\pi$, $g_A$, $M$, and $f_\pi$.  Thus, to
this order $\cpt$ makes {\it predictions} for Compton scattering.

The curves show that the correction from the $O(Q^3)$ terms gets
larger as $\omega$ is increased, as was to be expected. Indeed, while
at lower energies corrections are relatively small, in the 95 MeV
results the correction to the differential cross section from the
$O(Q^3)$ terms is of order 50\%, although the contribution of these
terms to the {\it amplitude} is of roughly the size one would expect
from the power-counting: about 25\%.  Nevertheless, it is clear, even
from these results, that this calculation must be performed to
$O(Q^4)$ before conclusions can be drawn about polarizabilities from
data at photon energies of order $m_\pi$. This is in accord with
similar convergence properties for the analogous calculation for
threshold pion photoproduction on the deuteron~\cite{silas2}.

We have also shown the Illinois data points at 69
MeV~\cite{lucas}. Statistical and systematic errors have been added in
quadrature.  The agreement of the $O(Q^3)$ calculation with the 69 MeV
data is very good, although only limited conclusions can be drawn,
given that there are only two data points, each with sizeable error
bars.

Although nominally the domain of validity of the Weinberg
formulation extends well beyond the threshold for pion production, the
power-counting fails at low energies well before the Thomson limit is
reached. By comparing $O(Q^4)$ and $O(Q^3)$ contributions, it
is straightforward to show that $\cpt$ breaks down when

\begin{equation}
{\frac{|\vpee\; |^2}{\omega M}}\sim 1.
\end{equation}
Here $\vpee$ is a typical nucleon momentum inside the deuteron and
$\omega$ is the photon energy.  Since our power-counting is predicated
on the assumption that all momenta are of order $m_\pi$, we find that
power-counting is valid in the region

\begin{equation}
\frac{m_\pi^2}{M}\ll Q \ll \lsc .
\end{equation}
Therefore, in the region $\omega \sim B$ the Weinberg power-counting
is not valid, since the external probe momentum flowing through the
nucleon lines is of order $Q^2/M$, rather than order $Q$. It is in
this region that the Compton low-energy theorems are
derived. Therefore our power-counting will not recover those
low-energy theorems. Of course the upper bound on the validity of the
effective theory should increase if the $\Delta$-resonance is included
as a fundamental degree of freedom.

In Ref.~\cite{martinetal} Compton scattering on the deuteron was
computed to the same order discussed here, one order beyond leading
non-vanishing order.  An advantage of KSW power-counting is that the
effective field theory moves smoothly between $Q < B$ and $Q > B$.
KSW power-counting is valid for nucleon momenta $Q< {\Lambda_{NN}}\sim
300$ MeV. Thus in the KSW formulation deuteron polarizabilities and
Compton scattering up to energies $\omega < {\Lambda_{NN}^2}/{M}\sim
90$ MeV can be discussed in the same framework.  Here we are
interested mostly in the region $\omega\sim m_\pi$, and so we regard
ourselves as being firmly in the second regime.  We stress that the
value of ${\Lambda_{NN}}$ is uncertain; it is conceivable that
${\Lambda_{NN}}\sim 500$ MeV in which case the range of the KSW
formulation would extend well beyond pion production threshold.

Comparing to the calculations of deuteron Compton scattering in the
KSW formulation of effective field theory~\cite{martinetal}, we see
that the result of Ref.  \cite{martinetal} is significantly lower than
those presented here at both 49 and 69 MeV. At 49 MeV (not shown here)
the agreement of Ref.~\cite{martinetal}'s calculation with the data is
better than ours.  Presumably this is partly because 49 MeV is at the
lower end of the domain of applicability of the Weinberg formulation.
At 69 MeV our calculation does a slightly better job of reproducing
the (two) data points available. The qualitative agreement among these
calculations is a reflection of the similarities of mechanisms
involved.  Ours is however the only calculation to incorporate the
full single-nucleon amplitude instead of its polarizability
approximation. (Note however that in Ref.~\cite{martinetal}, the first
corrections to the polarizability approximation of the pion graphs are
included and found to be very small.)  Our tendency to higher relative
cross sections in the backward directions is at least in part due to
this feature.

In order to test the sensitivity of our calculation to higher-order
effects we added a small piece of the $O(Q^4)$ amplitude for Compton
scattering off a single nucleon.  As one would expect, we find that
the cross section at 95 MeV is much more sensitive to these $O(Q^4)$
terms than the cross section at 49 MeV.  In our view, a full $O(Q^4)$
calculation in $\cpt$ is necessary if any attempt is to be made to
extract the neutron polarizability from the Saskatoon data within this
framework.


\section*{Acknowledgments}
\noindent I thank Veronique Bernard, Harry Lee, Man\'e Malheiro, Ulf
Mei{\ss}ner, Dan Phillips and Ubi van Kolck for enjoyable
collaborations and Ubi van Kolck for a critical reading of the
manuscript. This research was supported by DOE grant
DE-FG02-93ER-40762 (DOE/ER/40762-193, UMPP\#00-016).

\end{document}